# A New Iron Pnictide Oxide $(Fe_2As_2)(Ca_5(Mg,Ti)_4O_y)$ and a New Phase in Fe-As-Ca-Mg-Ti-O system

Hiraku Ogino[1,3][*], Yasuaki Shimizu[1,3], Naoto Kawaguchi[1,3], Kohji Kishio[1,3], Jun-ichi Shimoyama[1,3], Tetsuya Tohei[2], and Yuichi Ikuhara[2]

[1]Department of Applied Chemistry, The University of Tokyo, 7-3-1 Hongo, Bunkyo-ku, Tokyo 113-8656, Japan
[2]Institute of Engineering Innovation, The University of Tokyo, 2-11-16 Yayoi, Bunkyo-ku, Tokyo 113-8656, Japan
[3]JST-TRiP, Sanban-cho, Chiyoda-ku, Tokyo 102-0075, Japan
e-mail address: tuogino@mail.ecc.u-tokyo.ac.jp

## Abstract

A new layered iron arsenide oxide $(Fe_2As_2)(Ca_5(Mg,Ti)_4O_y)$ and its structural derivative were found in the Fe-As-Ca-Mg-Ti-O system. The crystal structure of $(Fe_2As_2)(Ca_5(Mg,Ti)_4O_y)$ is identical to that of $(Fe_2As_2)(Ca_5(Sc,Ti)_4O_y)$, which was reported in our previous study. The lattice constants of this compound are $a = 3.86(4)$ Å and $c = 41.05(2)$ Å. In addition, another phase with a thicker blocking layer was found. The structure of the compound and its derivative was tentatively assigned through STEM observation as $(Fe_2As_2)(Ca_8(Mg,Ti)_6O_y)$ with sextuple perovskite-type sheets divided by a rock salt layer. The interlayer Fe-Fe distance of this compound is ~30 Å. The compound and its derivative exhibited bulk superconductivity, as found from magnetization and resistivity measurements.

## Introduction

Since the discovery of high-$T_c$ superconductivity in LaFeAs(O,F)[1], several families of iron-based superconductors have been successively discovered. Thus far, new superconductors containing anti-fluorite iron pnictide or iron chalcogenide layers have been developed such as $BaFe_2As_2$[2], $LiFeAs$[3] and $FeSe$[4]. Recent discovery of new compounds in iron chalcogenide systems[5] indicates that there is still room for the development of new iron-based superconductors. Meanwhile, numerous layered iron pnictides composed of antifluorite-type iron pnictide layers and perovskite-type oxide layers have been recently found. Flexibility of the perovskite-based structure in such compounds has been investigated, and several structure types and constituent elements have been reported[6-17]. In our previous study, we discovered several homologous series of iron-based superconductors such as $(Fe_2As_2)(Ca_{n+1}(Sc,Ti)_nO_y)$ [$n$ = 3–5][14]. These compounds exhibit bulk superconductivity at a $T_c^{onset}$ of around 30–40 K without intentional carrier doping. Therefore, new compounds can be developed by increasing the thickness of the perovskite-type layer of already-known compounds such as $(Fe_2As_2)(Ca_4(Mg,Ti)_3O_y)$[15]. Moreover, systematic increase in the $T_c$ of $(Fe_2As_2)(Ca_{n+1}(Sc,Ti)_nO_y)$ and $(Fe_2As_2)(Ca_4(Mg,Ti)_3O_y)$ with a decrease in their $a$-axis length suggested that new compounds with higher $T_c$ can be realized by shortening their $a$-axis length, which may be accompanied by optimization of the local structure at the FeAs layer.

In this paper, we report a new iron pnictide superconductor, $(Fe_2As_2)(Ca_5(Mg,Ti)_4O_y)$. This compound is isostructural with $(Fe_2As_2)(Ca_5(Sc,Ti)_4O_y)$ having quadruple perovskite cells in the oxide layers between the FeAs layers. In addition, a new phase with a thicker blocking layer has been found. The structure of the compound is assigned as a stack of 6 layers of Ca-(Mg,Ti)-O perovskite sheets with one rock salt layer. Both compounds showed bulk superconductivity at a $T_c^{onset}$ of ~35 K in magnetization and ~40 K in resistivity measurements.

**Experimental**

Samples were synthesized through a solid-state reaction starting from FeAs(3N), CaO(3N), MgO(3N), Ti(3N), and $TiO_2$(3N). Nominal compositions were fixed according to the general formula, $(Fe_2As_2)(Ca_5(Mg_{1-x}Ti_x)_4O_y)$ ($0.6 \leq x \leq 0.9$, $9 \leq y \leq 11$). Since the starting reagents were sensitive to moisture in air, manipulation was carried out in a glove box filled with argon gas. Powder mixtures of FeAs, CaO, MgO, Ti, and $TiO_2$ were pelletized, sealed in evacuated quartz ampoules, and heated at 1000–1200°C for 72–100 hours. The constituent phases and lattice constants were analyzed through powder X-ray diffraction (XRD; MAC Science MXP-18) using Cu-$K_\alpha$ radiation, and intensity data were collected in the $2\theta$ range of 2°–80° in steps of 0.02°. Silicon powder was used as the internal standard. High-angle annular dark field (HAADF) images were obtained using a scanning transmission electron microscope (STEM; Cs-corrected JEM-2100F, JEOL). Magnetic susceptibility measurements were performed using a SQUID magnetometer (Quantum Design MPMS-XL5s). Electric resistivity was measured through the AC four-point-probe method using a Quantum Design PPMS.

**Result and discussion**

Figure 1(a) shows the powder XRD patterns of $(Fe_2As_2)(Ca_5(Mg_{1-x}Ti_x)_4O_{9.5})$ sintered at 1075°C for 100 h. A $(Fe_2As_2)(Ca_5(Mg,Ti)_4O_y)$ phase was formed as the main phase for the $x = 0.2$ sample, and there was a relatively large amount of impurities such as $CaTiO_3$, $CaFe_2As_2$, $MgO$, and $CaO$. It should be noted that no trace of $(Fe_2As_2)(Ca_4(Mg,Ti)_3O_y)$ phase was found in the XRD patterns. Samples starting from other compositions such as $x = 0.75$ or $x = 0.85$ contained larger amounts of impurities. Assuming that the titanium ion is tetravalent and that the oxygen sites are fully occupied, the ideal composition is considered to be $(Fe_2As_2)(Ca_5(Mg_{0.25}Ti_{0.75})_4O_{11})$. However, the $(Fe_2As_2)(Ca_5(Mg,Ti)_4O_y)$ phase was formed only from oxygen-poor starting compositions similar to $(Fe_2As_2)(Ca_{n+1}(M,Ti)_nO_y)(M = Sc,Al)$[14,16]. The optimal sintering temperature of 1075°C for the synthesis of $(Fe_2As_2)(Ca_5(Mg,Ti)_4O_y)$ phase was higher than that for the synthesis of $(Fe_2As_2)(Ca_4(Mg,Ti)_3O_y)$ phase, which is obtained by sintering at 1000°C as a main phase. A similar tendency was also found in the $(Fe_2As_2)(Ca_{n+1}(Sc,Ti)_nO_y)$ system: The optimal sintering temperatures were 1050, 1100 and 1200°C for compounds with $n = 3, 4$ and $5$, respectively. The crystal structure of $(Fe_2As_2)(Ca_5(Mg,Ti)_4O_y)$ consists of an alternate stacking of oxide layers with quadruple perovskite cells and antifluorite-type $Fe_2As_2$ layers. The space group of this structure is *I4/mmm*, and the lattice constants were determined to be $a = 3.86(4)$ Å and $c = 41.05(2)$ Å. The *a*-axis of this structure is shorter than that of $(Fe_2As_2)(Ca_4(Mg,Ti)_3O_y)$ (3.877 Å)[15]. The difference between the *a*-axis values of these compounds is probably due to the higher titanium content in $(Fe_2As_2)(Ca_5(Mg,Ti)_4O_y)$ than in $(Fe_2As_2)(Ca_4(Mg,Ti)_3O_y)$ phase; this higher titanium content is because the ionic radius of $Ti^{4+}$ (0.605 Å) is smaller than that of $Mg^{2+}$ (0.72 Å). It should be noted that apparent changes in the lattice constants were not found among samples starting from different compositions and sintered under different conditions. This fact suggests that chemical compositions of the resulting $(Fe_2As_2)(Ca_5(Mg,Ti)_4O_y)$ phases are similar to each other.

In addition to the $(Fe_2As_2)(Ca_5(Mg,Ti)_4O_y)$ phase, a different phase with a longer stacking pattern was found in the samples sintered at high temperatures above 1100°C. The XRD pattern of $(Fe_2As_2)(Ca_5(Mg_{0.25}Ti_{0.75})_4O_{9.5})$ sintered at 1125°C is shown in Fig. 2. As shown in the inset, a sharp peak was observed at 2.9°, corresponding to a layer spacing of ~30 Å. Although the sample contained large amounts of impurities such as $CaFe_2As_2$, $CaTiO_3$, and MgO, most of the peaks were indexed by a primitive tetragonal cell with $a \sim 3.86$ Å and $c \sim 30$ Å. Figure 3(a) and (b) shows HAADF-STEM images and electron diffraction (ED) patterns taken from the [100] direction of the sample $(Fe_2As_2)(Ca_5(Mg_{0.25}Ti_{0.75})_4O_{9.5})$ sintered at 1125°C. The FeAs layers were imaged as arrays of bright dumbbell-like contrasts, and the cation sites of the perovskite block layers in between were observed as less bright spots. It was found that the structure of the compound consists of an alternate stacking of anti fluorite $Fe_2As_2$ layer and a very thick perovskite-type layer. The perovskite-type layer of the compound was composed of sextuple perovskite-type sheets divided by a rock salt layer, as shown in Fig. 3(c). The chemical formula of the structure is $(Fe_2As_2)(Ca_8(Mg,Ti)_6O_y)$. The ED pattern of the compound indicated a tetragonal cell with a $c/a$ of ~8.0. The value was coincident with that estimated through XRD analysis, which was ~7.8. Similar to $(Fe_2As_2)(Ca_{n+1}(Sc,Ti)_nO_y)$, a weak superstructure along the $a$-axis was observed in the ED pattern.

We have tried to reduce the amount of impurities by changing the starting compositions, synthesis conditions, and synthesis procedures; however, none of the attempts has been successful so far. Presently, the reason for the large difference between the starting composition and the ideal chemical formula is not clear. Unlike $(Fe_2As_2)(Ca_{n+1}(Sc,Ti)_nO_y)$ and $(Fe_2As_2)(Ca_{n+1}(Al,Ti)_nO_y)$, stacking faults along the $c$-axis direction were often observed. Fig. 3(d) shows a typical stacking fault observed in the same sample. The structure at the fault was composed of triple-perovskite cells/rock salt layer/double-perovskite cells. This structure indicates that another compound with the chemical formula $(Fe_2As_2)(Ca_7(Mg,Ti)_5O_y)$ may have been synthesized by optimizing the synthesis conditions. Other kinds of stacking faults were sometimes observed in different grains of the same sample. The formation of stacking faults in this system can be explained by the difference between the $Mg^{2+}/Ti^{4+}$ ratios of the regular layer and the stacking fault layer. The ideal valence at the (Mg,Ti) site in $(Fe_2As_2)(Ca_8(Mg,Ti)_6O_y)$ is calculated from the chemical formula($(Fe_2As_2)(Ca_8(Mg,Ti)_6O_{18})$) as +3.67, while that in $(Fe_2As_2)(Ca_7(Mg,Ti)_5O_y)$ is +3.60. To adjust the cation valence to $M^{+3.6}$ and $M^{+3.67}$ by mixing Mg and Ti, the expected $Mg^{2+}/Ti^{4+}$ ratios are 0.25 and 0.2 and the mean ionic radii are 0.628 and 0.624 Å (coordination number VI), respectively. On the other hand, those of $Sc^{3+}/Ti^{4+}$ are

0.67(0.661 Å) and 0.5(0.652 Å), respectively. Because of close ratio of the cations and ionic radii in the (Mg,Ti) system, conditions of $(Fe_2As_2)(Ca_8(Mg,Ti)_6O_y)$ and $(Fe_2As_2)(Ca_7(Mg,Ti)_5O_y)$ phases might be very similar. The interlayer Fe-Fe distance in the $(Fe_2As_2)(Ca_8(Mg,Ti)_6O_y)$ phase is ~30 Å, which is the longest among the layered iron pnictides ever reported. Such an extremely anisotropic crystal structure is very rare even in simple oxides.

The temperature dependences of zero-field-cooled (ZFC) and field-cooled (FC) magnetization of $(Fe_2As_2)(Ca_5(Mg,Ti)_4O_y)$ and $(Fe_2As_2)(Ca_8(Mg,Ti)_6O_y)$ measured under 1 Oe are shown in Fig. 4. Although the demagnetization effect for these samples was roughly estimated to be ~20%, it was not accounted for in the magnetic susceptibility shown in this figure, owing to the complicated shapes and porosity of the samples. Both compounds showed diamagnetism due to superconductivity at ~40 K, and the superconducting volume fraction estimated from ZFC magnetization at 2K for the compounds was ~80% and 100%, respectively. The reversible region, where the ZFC and FC magnetization curves overlap each other, was observed down to 33 and 28 K, suggesting a granular nature with weak pinning similar to that observed in other layered iron pnictides having perovskite-type oxide blocking layers[11, 14-16]. These compounds showed superconductivity without intentional carrier doping, but further investigation is required to clarify the carrier doping mechanisms of the compounds.

The magnetic hysteresis loops of the $(Fe_2As_2)(Ca_5(Mg,Ti)_4O_y)$ sample were measured at 5, 7, and 10 K. A weak ferromagnetic component, which was probably owing to the presence of impurities, was also observed. Small hysteresis loop widths at each temperature indicate weak grain connection and electromagnetic granularity of the sample. Although the $T_c$ value observed in magnetization measurements was above 30 K, $J_c$ increased substantially with decreasing temperature from 10 to 5 K. Under the assumption that the hysteresis loop width mainly originate from intra-grain $J_c$($J_c^i$) of each crystal, $J_c^i$ is estimated to be of the order of $10^5$ A/cm$^2$ at 5 K in low fields, based on the Bean model, according to which $J_c = 30\Delta M/d$, where $d$ is the mean crystal size, 20 μm.

Figure 6 shows the temperature dependence of resistivity for $(Fe_2As_2)(Ca_5(Mg,Ti)_4O_y)$ measured under various magnetic fields. Metallic behavior was observed in the normal-state resistivity. The resistivity of $(Fe_2As_2)(Ca_5(Mg,Ti)_4O_y)$ steeply reduced at ~40 K, and zero resistance was achieved at 35K for $H = 0$. The largely and systematically broadened resistivity transitions obtained by increasing the external fields might be attributed to the poor flux pinning originating from the thick blocking layer as well as the weak coupling between grains. The irreversibility temperature, determined by 1/1000 of normal state resistivity was 20, 17, 16, and 15 K for 1, 3, 5, and 9 T, respectively. Superconducting transition was also observed in the resistivity curve of $(Fe_2As_2)(Ca_8(Mg,Ti)_6O_y)$. As in the case of $(Fe_2As_2)(Ca_5(Mg,Ti)_4O_y)$, the $T_c^{onset}$ of the sample was around 40 K and zero resistance was achieved at 22 K. It was concluded from the temperature dependence of resistivity that this compound was metallic and almost proportional to temperature.

The values of $a$-axis lengths and $T_c^{onset}$ in resistivity curves of iron arsenides with perovskite-type blocking layers are summarized in Fig. 7. The $T_c$ systematically increases with the values of the $a$-axis length, and maximum $T_c$ is achieved at around 3.88 Å. However, further increase in the $a$-axis length decreases the $T_c$. The exception at ~3.92 Å is $T_c$ of $(Fe_2As_2)(Sr_4Cr_2O_6)$, in which Cr substitution to Fe site is reported[19]. Because the $a$-axis length is correlated with $h_{pn}$[19], the observed variation in $T_c$ with the $a$-axis length might be related to the fact that there is a maximum of $T_c$ with increasing $h_{pn}$[20]. On the other hand, superconductivity is not observed for the compounds with longer $a$-axis length, such as $(Fe_2As_2)(AE_3Sc_2O_5)$ and $(Fe_2As_2)(AE_4Sc_2O_6)$ ($AE$ = Sr, Ba)[6,8,10,13], to the best of our knowledge. The presence of magnetic ordering of the $Fe_2As_2$ layer in these compounds is still a matter of debate[21,22]. The absence of superconductivity in $(Fe_2As_2)(Sr_4Sc_2O_6)$ cannot be explained by $h_{pn}$, because the $h_{pn}$ of the compound is smaller than that of LaFeAsO[1,18]; therefore, the superconductivity might be related to the direct Fe-Fe intralayer distances.

**Conclusions**

A new layered iron pnictide oxide $(Fe_2As_2)(Ca_5(Mg,Ti)_4O_y)$, which is isostructural with $(Fe_2As_2)(Ca_5(Sc,Ti)_4O_y)$, was synthesized. In addition, a new phase with longer stacking pattern was observed. The structure of the compound is assigned as $(Fe_2As_2)(Ca_8(Mg,Ti)_6O_y)$. This compound exhibited substantial diamagnetism below ~35 K, suggesting bulk superconductivity at low temperatures without intentional carrier doping. On resistivity measurement under $H = 0$, this compound showed metallic behavior with a $T_c^{onset}$ of ~40 K.

Our present results suggest that the layered iron pnictides with perovskite-type blocking layers contain a large variety of crystal structures and constituent elements at the oxide layer. Such a large variety will contribute to the understanding of the mechanism of iron-based superconductors.

**Acknowledgements**

The authors thank Mr. K. Ushiyama for his assistance in conducting the experiments at a primary stage and Mr. S. Sato for his fruitful discussions. This work was partly supported by a Grant-in-Aid for Young Scientists (B) No. 21750187, 2009, from the Ministry of Education, Culture, Sports, Science and Technology (MEXT), Japan. A part of this work was conducted at the Center for Nano Lithography & Analysis, The University of Tokyo, supported by MEXT, Japan

**Figure captions**

Figure 1. Powder XRD patterns of $(Fe_2As_2)(Ca_5(Mg_{1-x},Ti_x)_4O_{9.5})(0.75 \leq x \leq 0.85)$ sintered at 1075°C(a) and crystal structure of $(Fe_2As_2)(Ca_5(Mg,Ti)_4O_y)$ (b).

Figure 2. Powder XRD pattern of $(Fe_2As_2)(Ca_5(Mg_{0.25},Ti_{0.75})_4O_{9.5})$ sintered at 1125°C. XRD pattern from 2°–6° are shown in the inset.

Figure 3. HAADF-STEM image of $(Fe_2As_2)(Ca_5(Mg_{0.25},Ti_{0.75})_4O_{9.5})$ sample sintered at 1125°C viewed from the [100] direction (a). The circles in the inset indicate the arsenic, iron, calcium, and aluminum/titanium cation columns. ED patterns of the sample (b). Crystal structure of $(Fe_2As_2)(Ca_8(Mg,Ti)_6O_y)$ (c). Typical stacking fault observed in the same sample (d) The position of the fault is indicated by the arrow. A magnification of the stacking fault image viewed from the [100] direction (e).

Figure 4. Temperature dependence of ZFC and FC magnetization curves of the $(Fe_2As_2)(Ca_5(Mg,Ti)_4O_y)$ bulk sample with a nominal composition of $(Fe_2As_2)(Ca_5(Mg_{0.2}Ti_{0.8})_4O_{9.5})$ sintered at 1075°C (closed circle) and the $(Fe_2As_2)(Ca_5(Mg,Ti)_4O_y)$ bulk sample with a nominal composition of $(Fe_2As_2)(Ca_5(Mg_{0.25}Ti_{0.75})_4O_{9.5})$ sintered at 1125°C (open triangle) measured under 1 Oe. A magnification of the magnetization curves is shown in the inset.

Figure 5. Magnetization hysteresis loop of the polycrystalline bulk of $(Fe_2As_2)(Ca_5(Mg,Ti)_4O_y)$ at 5, 7, and 10 K.

Figure 6. Temperature dependence of resistivity under various magnetic fields of the $(Fe_2As_2)(Ca_5(Mg,Ti)_4O_y)$ bulk sample with a nominal composition of

($Fe_2As_2$)($Ca_5(Mg_{0.2}Ti_{0.8})_4O_{9.5}$) (a). Irreversible temperatures under each magnetic field are indicated. A resistivity curve up to 300 K for $H = 0$ is shown in the inset. Temperature dependence of resistivity of the ($Fe_2As_2$)($Ca_8(Mg,Ti)_6O_y$) bulk sample with a nominal composition of ($Fe_2As_2$)($Ca_5(Mg_{0.25}Ti_{0.75})_4O_{9.5}$) under various magnetic fields (b).

Figure 7. Relationship between the values of $T_c^{onset}$ and *a*-axis length of iron pnictides with perovskite-type blocking layers. $T_c^{onset}$ is determined by the onset of the resistivity curve. *M* cations in the Fe-As-AE-*M*-O system are indicated in the figure.

Figure 1

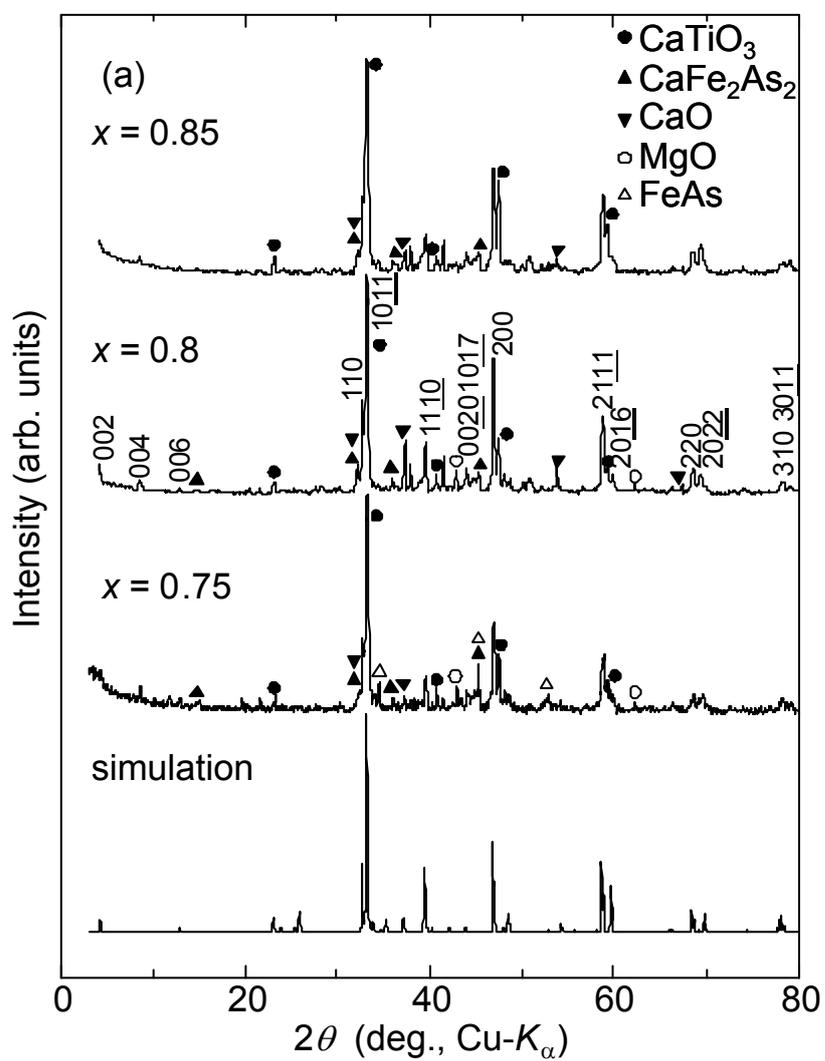

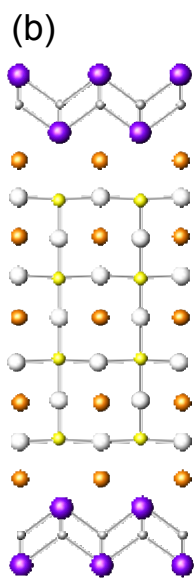

Figure 2

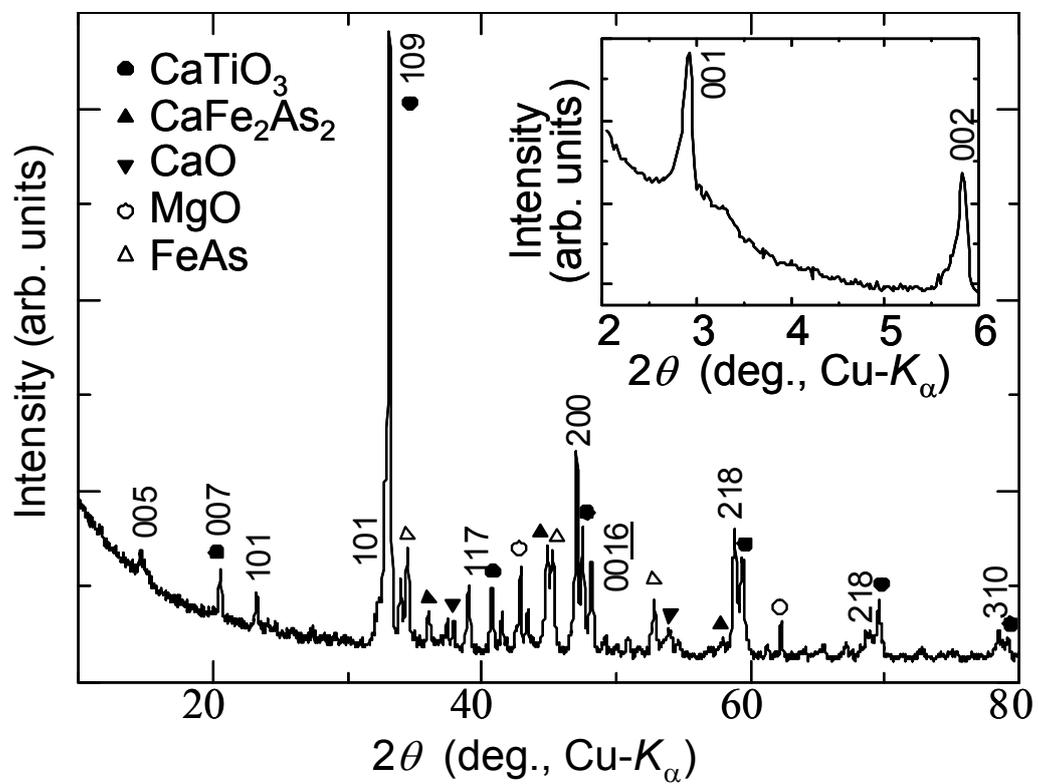

Figure 3

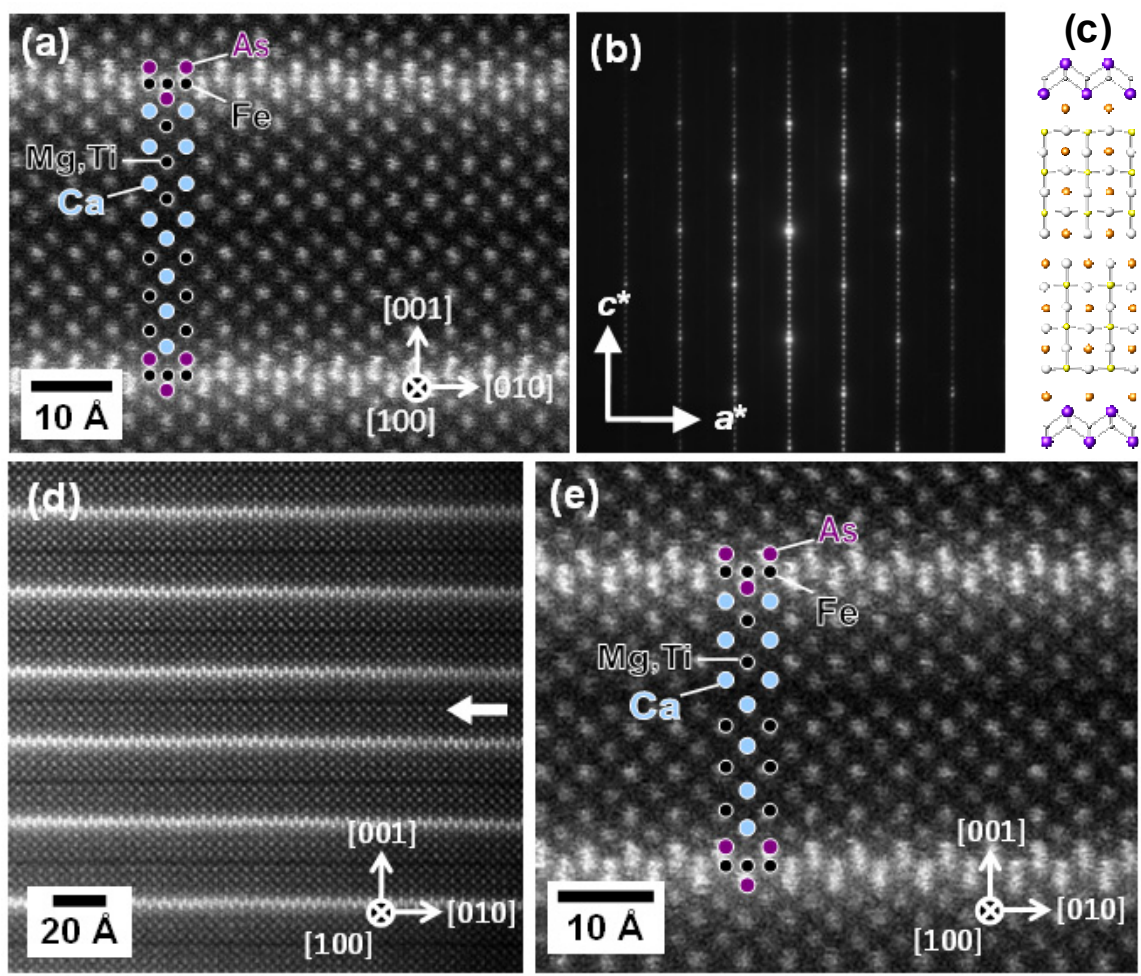

Figure 4

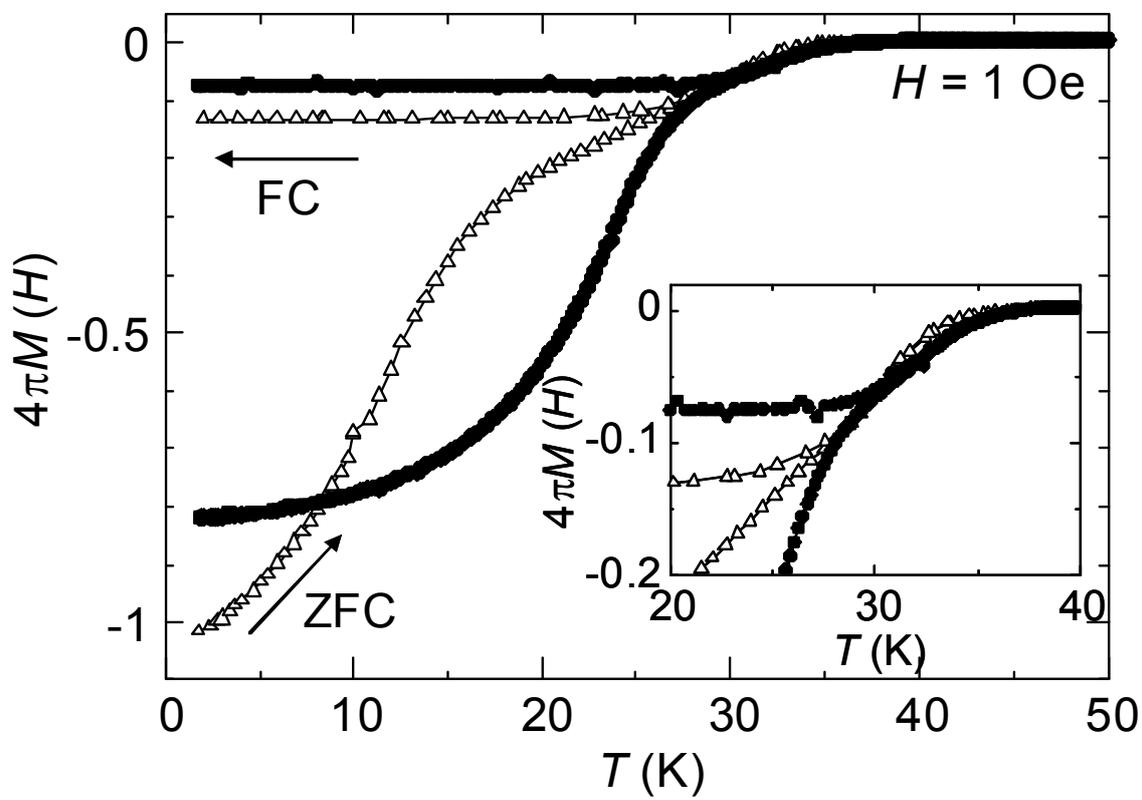

Figure 5

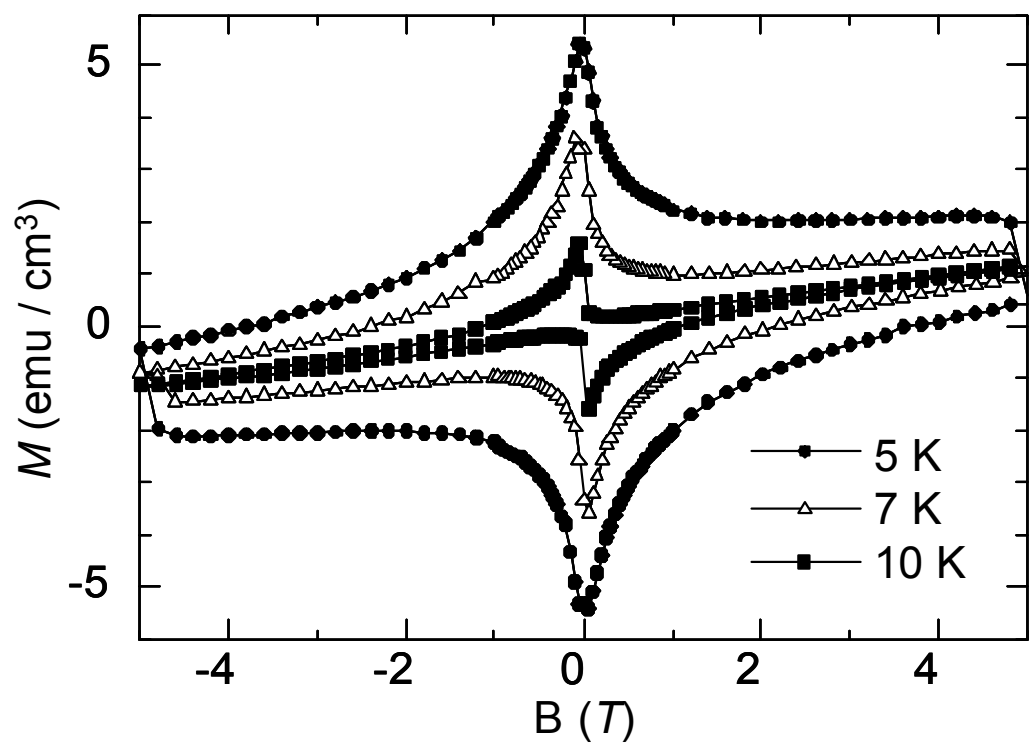

Figure 6

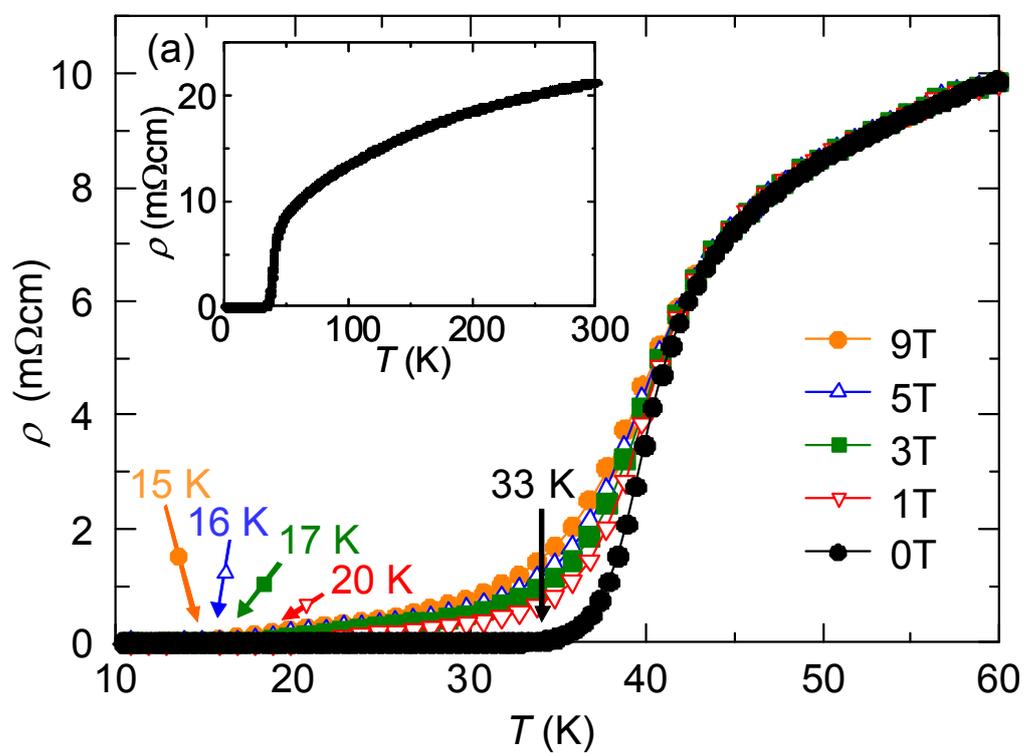

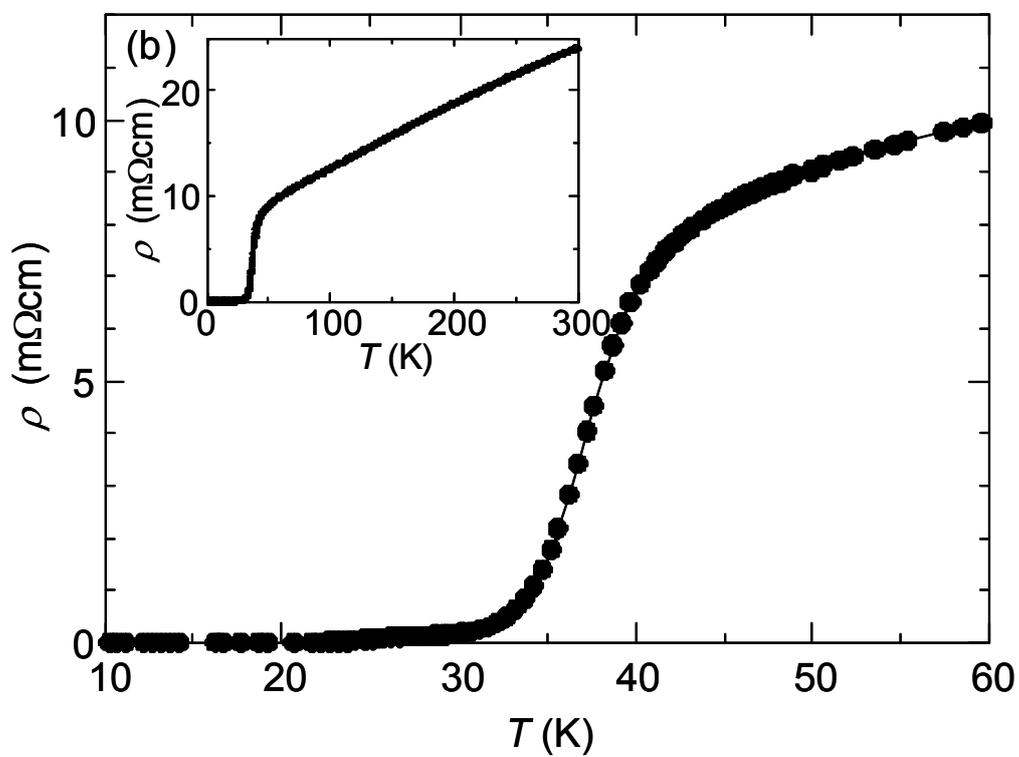

Figure 7

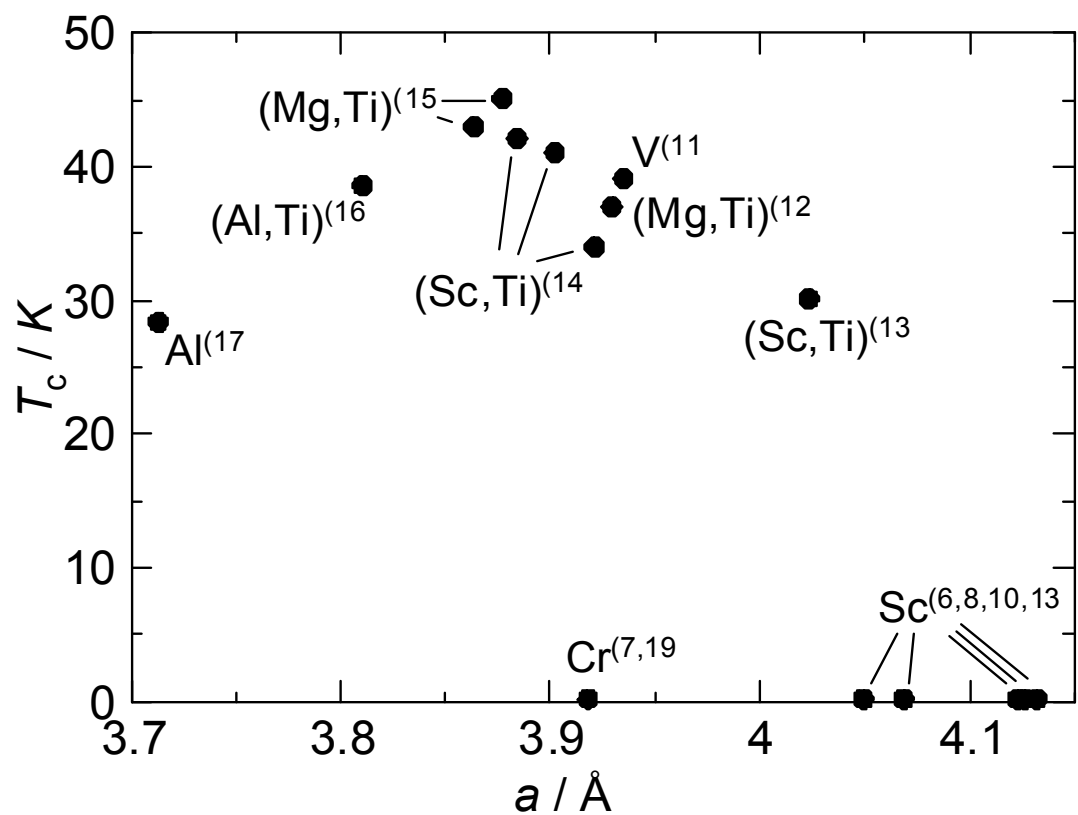